\documentclass[letterpaper,american,british,reprint, aps,superscriptaddress]{revtex4-1}
\usepackage[T1]{fontenc}
\usepackage[utf8]{inputenc}
\usepackage{amsmath}
\usepackage{amssymb}
\usepackage{stmaryrd}
\usepackage{graphicx}

\makeatletter

\pdfpageheight\paperheight
\pdfpagewidth\paperwidth

\providecommand{\tabularnewline}{\\}

\makeatother

\usepackage[usenames,dvipsnames]{xcolor}
\usepackage[pdftex, hidelinks]{hyperref}

\newcommand\myshade{85}
\colorlet{mylinkcolor}{violet}
\colorlet{mycitecolor}{YellowOrange}
\colorlet{myurlcolor}{Aquamarine}
\hypersetup{
	linkcolor  = mylinkcolor!\myshade!black,
	citecolor  = mycitecolor!\myshade!black,
	urlcolor   = myurlcolor!\myshade!black,
	colorlinks = true,
}

\usepackage{babel}
\begin{document}
\title{Entanglement distribution with wavevector-multiplexed quantum memory}
\author{Michał Lipka}
\affiliation{Centre for Quantum Optical Technologies, Centre of New Technologies,
University of Warsaw, Banacha 2c, 02-097 Warsaw, Poland}
\affiliation{Faculty of Physics, University of Warsaw, Pasteura 5, 02-093 Warsaw,
Poland}
\author{Mateusz Mazelanik}
\affiliation{Centre for Quantum Optical Technologies, Centre of New Technologies,
University of Warsaw, Banacha 2c, 02-097 Warsaw, Poland}
\affiliation{Faculty of Physics, University of Warsaw, Pasteura 5, 02-093 Warsaw,
Poland}
\author{Michał Parniak}
\email{parniak@nbi.ku.dk}

\affiliation{Centre for Quantum Optical Technologies, Centre of New Technologies,
University of Warsaw, Banacha 2c, 02-097 Warsaw, Poland}
\affiliation{Niels Bohr Institute, University of Copenhagen, Blegdamsvej 17, DK-2100
Copenhagen, Denmark}
\begin{abstract}
Feasible distribution of quantum entanglement over long distances
remains a fundamental step towards quantum secure communication and
quantum network implementations. Quantum repeater nodes based on quantum
memories promise to overcome exponential signal decay inherent to
optical implementations of quantum communication. While performance
of current quantum memories hinders their practical application, multimode
solutions with multiplexing can offer tremendous increase in entanglement
distribution rates. We propose to use a wavevector-multiplexed atomic
quantum memory (WV-MUX-QM) as a fundamental block of a multiplexed
quantum repeater architecture. We show the WV-MUX-QM platform to provide
quasi-deterministic entanglement generation over extended distances,
mitigating the fundamental issue of optical loss even with currently
available quantum memory devices, and exceeding performance of repeaterless
solutions as well as other repeater-based protocols such as temporal
multiplexing. We establish the entangled-bit (ebit) rate per number
of employed nodes as a practical figure of merit reflecting the cost-efficiency
of larger inter-node distances.
\end{abstract}
\maketitle

\section{Introduction}

Entanglement is an essential resource for the most promising quantum
information protocols \citep{Sangouard2011,Wehner2018} enabling,
among others, secure quantum communication \citep{Pirandola2015,Pirandola2019,Zhang2017}.
The optical implementations of such protocols face the exponential
transmission losses inherent to photonic systems and greatly limiting
the feasible distance at which high fidelity entangled states can
be distributed. To overcome this obstacle, noise-tolerant quantum
repeaters have been proposed for entanglement connection and purification
over shorter elementary lengths \citep{Briegel1998}.

Experimentally promising quantum repeater architectures involve linear
optics, quantum memories and single-photon detection \citep{Yuan2008,Zhao2007};
however, currently available memory lifetimes as well as retrieval
and single-photon detection efficiencies limit the feasibility of
such repeaters at practical distances of a few hundred km \citep{Gisin2015,Yu2020}.
Multimode architectures have been proposed as solutions to this problem
\citep{Collins2007,Dam2017,Simon2007}, which lead to an ongoing effort
in experimental realizations of such systems, especially involving
multiplexing capabilities. While an $M$-mode platform in parallel
operation increases the entanglement distribution rate $M$-fold,
multiplexing may lead to $\mathcal{O}(M^{2N})$-fold increase with
$N$ denoting the number of repeater nodes \citep{Collins2007}. Hitherto
multimode systems demonstrated in the context of quantum repeaters
involved at most tens of modes \citep{Simon2010,Chang2019,Sinclair2014,Wen2019,Tian2017,Li2019a}
and mainly focused on temporal multiplexing. As an alternative to
temporal, spectral or spatial micro-ensemble modes \citep{Pu2017},
a highly-multimode wavevector multiplexed quantum memory (WV-MUX-QM)
has been recently demonstrated \citep{Parniak2017} along with flexible
in-memory processing capabilities \citep{Parniak2019,Mazelanik2019,Lipka2019}.

Here we evaluate the feasibility of previously-demonstrated WV-MUX-QM,
which was based on a high-optical-depth cold atomic ensemble, as a
quantum repeater platform. Remaining in the constraints of current
technology, we propose a multiplexing protocol combining experimentally
demonstrated components to provide quasi-deterministic entanglement
generation over $150\;\text{km}$.

We analyse the performance of the novel platform in the recently proposed
semihierarchical quantum repeater architecture \citep{Liu2017} as
well as an ahierarchical architecture and compare the performance
of wavevector multiplexing with state-of-the art temporal multimode
and long-lifetime single-mode platforms. Finally, we identify the
fundamental limitations of the WV-MUX-QM platform. Our results are
particularly significant in the light of recent advances and effort
in developing practically feasible multi-mode quantum communication
systems employing multicore fibres \citep{Richardson2013,Ding2017}
or free-space transmission \citep{Liao2017}.

\subsection*{Quantum repeaters – DLCZ protocol}

As proposed by Duan, Lukin, Cirac and Zoller (DLCZ) in their seminal
paper \citep{Duan2001}, atomic ensemble-based quantum memories and
linear optical operations combined with single-photon detection can
be employed to transfer entanglement between distant parties. In a
simple scenario of the DLCZ protocol, two parties – Alice (A) and
Bob (B) – are separated by a distance $L$ and would like to share
a high-fidelity entangled photonic state for further use in quantum
communication protocols such as Quantum Key Distribution (QKD). If
the optical losses are too large over the distance $L$, we can imagine
$N$ equidistant parties separated by $L_{0}=L/(N-1)$. Each party
($A_{i}$) has two atomic ensembles and independently generates entanglement
between one of the ensembles and one of her neighbours ($A_{i-1}$,
$A_{i+1}$). Then, hierarchically, the parties read their two ensembles
and, conditioned on a successful detection of a single read-out photon,
extend the entanglement to their neighbours ($A_{i-1}\leftrightarrow A_{i}\leftrightarrow A_{i+1}\;\rightarrow\;A_{i-1}\leftrightarrow A_{i+1}$).

The DLCZ protocol, being based on single-photon interference, requires
sub-wavelength phase stability over tens-of-km inter-node distances
$L_{0}$, limiting its practical feasibility. Furthermore, in a DLCZ-based
network the distributed entangled state contains a vacuum component
which grows with every entanglement swapping stage. As a solution
to those issues, two-photon interference-based protocols have been
proposed \citep{Zhao2007,Chen2007} and experimentally demonstrated
\citep{Yuan2008,Yu2020}. In this work we build on the two-photon
protocols \citep{Zhao2007,Chen2007} inherently robust to optical
phase stability. In particular, while we design the entanglement generation
(ENG) scheme specifically for WV-MUX-QM platform, the entanglement
connection (ENC), which takes place after multiplexing (MUX), remains
the same as in standard two-photon protocols.

The basic idea of two-photon protocols is to employ polarization entangled
pairs of photons instead of occupation number entangled state with
a single photon. In a basic scenario the $i$-th node $A_{i}$ has
a pair of memories $A_{i}^{(H,L)}$, $A_{i}^{(V,L)}$ for ENG with
$A_{i-1}$ and another pair $A_{i}^{(H,R)}$, $A_{i}^{(V,R)}$ for
$A_{i+1}$. Polarization of photons emitted from the memories is transformed
so that the superscripts $H$ and $V$ correspond to horizontal and
vertical polarization of the photons, respectively. Similarly to the
DLCZ protocol, photons from $A_{i}$ and $A_{i+1}$ are sent to the
midway station and interfered on a beamsplitter; however, now the
interference and single-photon detection is separate for each polarization.
For ENC both $H$ and $V$ memories are readout and the state of the
distant pairs of memories is projected onto a maximally entangled
Bell state by measuring the read-out photons and post-selecting outcomes.

\section{Results}

\subsection*{Wavevector-multiplexed quantum repeater}

\subsubsection*{Multimode quantum memory}

\begin{figure}
\includegraphics[width=1\columnwidth]{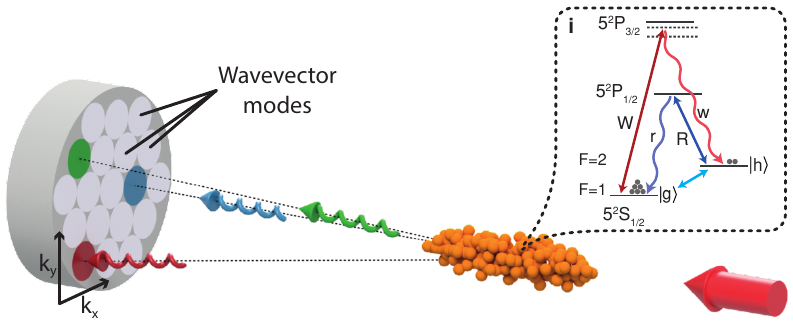}

\caption{With wavevector multiplexing, each mode of such an atomic memory corresponds
to a different angle of emission (or wavevector) of write-out (read-out)
photons. \textbf{i} Atomic levels in a lambda configuration for a
light-atom interface to the wavevector multiplexed quantum memory
based on cold rubidium-87 atoms. W - write beam, R - read beam, w
- write-out photon, r - read-out photon. For technical reasons our experiments presented in \cite{Parniak2017} used different atomic transitions for writing and reading. \label{fig:lambda_modes}}
\end{figure}

The WV-MUX-QM platform is based on an atomic quantum memory. For specificity
we shall consider rubidium-87 atoms \citep{Parniak2017} cooled in
a magneto-optical trap (MOT) via polarization gradient cooling and
trapped in a dipole trap \citep{Grimm2000} reaching temperatures
of $1\;\mu\text{K}$. The memory operates via a light-atom interface
based on an off-resonant spontaneous Raman scattering. A lambda configuration
of atomic levels is involved, as depicted in the inset of Fig. \ref{fig:lambda_modes}.
Importantly, the selected atomic transitions should be clock transitions
robust to external magnetic fields \citep{Zhao2009b}. The atoms are
initially prepared in the $|g\rangle$ level. A strong write (W) beam
off-resonant with $|g\rangle\rightarrow|e\rangle$ transition generates
a two-mode squeezed state of scattered write-out (Stokes) photons
and collective atomic excitations (spin-waves, coherence between $|g\rangle$
and $|h\rangle$ levels). Let us denote by $\hat{a}_{{\bf k_{w}}}^{\dagger}$
the creation operator for write-out photon with a wavevector $\mathbf{k_{w}}$
determining the scattering angle and the memory mode, as depicted
in Fig. \ref{fig:lambda_modes}. Denoting creation operator for a
single spin-wave with wavevector $\mathbf{K}$ by $\hat{S}_{{\bf K}}^{\dagger}$,
the generated state in a single pair of memory and photonic modes
is given by:
\begin{equation}
|\psi\rangle=\sum_{j=0}(\sqrt{\chi}\hat{S}_{{\bf K}}^{\dagger}\hat{a}_{{\bf k_{w}}}^{\dagger})^{j}|\text{vac}\rangle,\label{eq:inmem}
\end{equation}
where $\chi$ gives the probability of generating a single pair of
photon-spin-wave excitations. After a programmable delay, the spin-waves
can be converted to read-out photons with a strong resonant read (R)
beam $\hat{S}_{{\bf K}}^{\dagger}\rightarrow\hat{a}_{{\bf k_{r}}}^{\dagger}$.

Write and read processes conserve momentum and energy which results
in correlated momenta or wavevectors (or scattering angles) of write-out
and read-out photons. While, in general, phase-matching conditions
must be taken into consideration, the most versatile write/read beams
configuration employs counter-propagating beams at angle of around
$2^{\circ}$ to the longitudinal axis of MOT \citep{Parniak2017}.
Importantly, with such a choice of write and read beams' wavevectors,
the write-out photon at $\mathbf{k_{w}}$ heralds further read-out
to result in a read-out photon at $\mathbf{k_{r}\approx-\mathbf{k_{w}}}$,
which is fundamental for wavevector multiplexing.

\begin{figure}
\includegraphics[width=1\columnwidth]{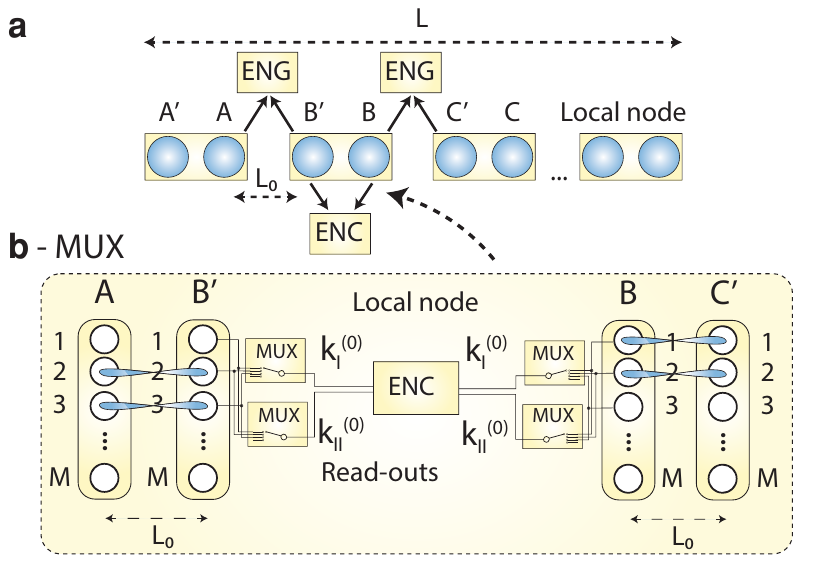}

\caption{\textbf{a} Entanglement distribution with an ahierarchical quantum
repeater architecture and employing atomic wavevector multiplexed
quantum memories (WV-MUX-QM). Each node consists of two WV-MUX-QM
(e.g. $A'$,$A$). Intermediate nodes are located $L_{0}$ apart and
send the write-out photons generated in one of their atomic ensembles
to a midway station in order to generate entanglement (ENG) between
the distant ensembles (e.g. $A$ and $B'$).\textbf{ b} Multiplexing
idea – to entangle ensembles $A$ and $C'$ in entanglement connection
stage (ENC) after successful ENG between $A\ \&\ B'$ and $B\ \&\ C'$,
the read-out photons from $B$ and $B'$ are multiplexed to pre-selected
modes $\mathbf{\mathbf{k_{I}^{(0)}}}$, $\mathbf{k_{II}^{(0)}}$ with
a reconfigurable MUX and using the which-mode information from ENG
stages. \label{fig:idea}}
\end{figure}

\begin{figure*}
\includegraphics[width=1\textwidth]{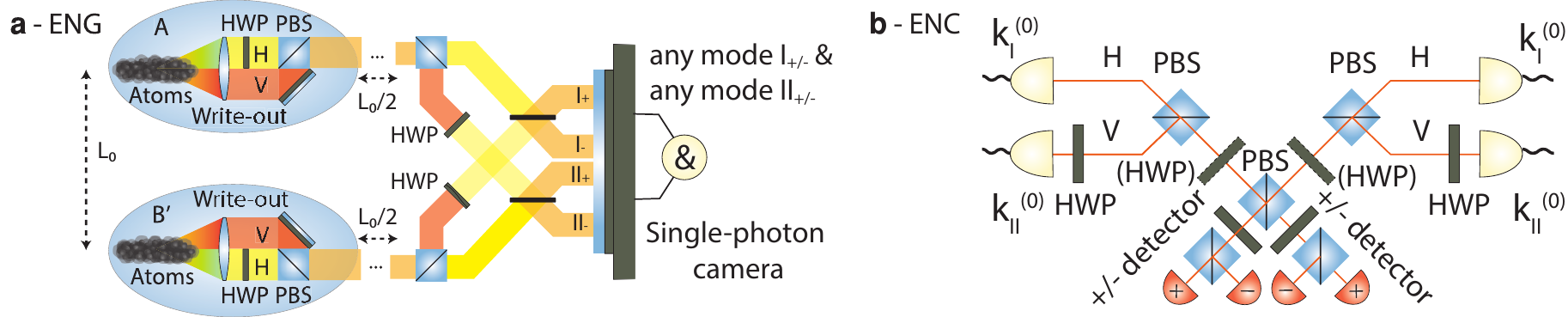}

\caption{\textbf{a} The WV-MUX-QM involves an atomic ensemble optically interfaced
via spontaneous Raman scattering which generates pairs of a write-out
photon and a collective atomic-excitation – a spin-wave. In the far-field
of the atomic ensemble the wavevector (angular) modes of write-out
photons are divided into two groups with imposed horizontal ($H$)
or vertical ($V$) polarization, and the groups are super-imposed.
Entanglement generation (ENG) between two distant WV-MUX-QM memories
– $A'$ and $B$ involves multimode transmission of write-out photons
to a midway station. Due to spatially resolved single-photon detection
(single-photon camera), any of $M$ modes from $A'$ can be entangled
with any of $M$ modes from $B$ giving $M^{2}$ possibilities and
facilitating quasi-deterministic ENG even with substantial losses.
The ENG is heralded by a detection of a coincidence between a photon
in regions $\mathbf{I_{\pm}}$ and in regions $\mathbf{II_{\pm}}$.\textbf{
b} Entanglement connection (ENC) stage takes place at a local node.
The memories are read and read-out photons are multiplexed to pre-selected
optical modes $\mathbf{\mathbf{k_{I}^{(0)}}}$, $\mathbf{k_{II}^{(0)}}$.
Upon a coincidence detection between different $\pm$ detectors, the
optical setup projects the spin-wave state of ensembles $A$ and $C'$
onto a maximally entangled Bell state. PBS - polarizing beamsplitter,
HWP - half-wave plate set at $45^{\circ}$, (HWP) - optional HWP for
first-stage ENC.\label{fig:engenc}}
\end{figure*}

\subsubsection*{Entanglement distribution}

An ahierarchical quantum repeater architecture has been depicted in
Fig. \ref{fig:idea}a. Each node contains two WV-MUX-QMs. Distant
memories perform entanglement generation (ENG) by sending write-out
photons to a central midway station. To distribute the entanglement,
memories are read at each node and the read-out photonic modes are
multiplexed to pre-selected modes, as depicted in Fig. \ref{fig:idea}b,
to further undergo entanglement connection (ENC).

\subsubsection*{Entanglement generation (ENG)}

The first operation in a quantum repeater link is the generation of
entanglement (ENG) between each pair of quantum memories separated
by an elementary distance $L_{0}=L/(N-1)$. The memories probabilistically
generate entangled pairs of atomic excitations and single-photons
across $M$ modes of their ensembles. The generated write-out photons
are sent to a central mid-way station (CMS) via multimode channels.
We note here that there is no inter-mode phase stability requirement,
rendering either free space or multimode fibres a viable multimode
channel. On the other hand, a space-to-time conversion \citep{Boffi1996}
could be used to map the wavevector modes to time bins and thus enabling
use of single-mode channels for ENG. The photons generated in each
ensemble are divided by their emission angles into two groups with
imposed orthogonal polarizations – vertical ($V$) and horizontal
($H$), as depicted in Fig. \ref{fig:engenc}a. Before transmission,
the two groups are super-imposed on each-other to ensure that each
pair of $H,V$ modes is transmitted through a single channel acquiring
the same optical phase or a deterministic phase difference.

Let us consider two ensembles $A$, $B'$ located in separate nodes.
Horizontal (vertical) polarization from $A$ is superimposed onto
vertical (horizontal) polarization from $B'$, erasing the which-node
information and resulting in four regions containing $M$ modes each.
The regions $\mathbf{I}_{\pm}$ ($\mathbf{II}_{\pm}$) observe $V$
($H$) polarization from $B'$ and $H$ ($V$) polarization from $A$.
A single-photon sensitive camera \citep{Lipka2018} observes the regions
and registers coincidences between any of modes in $\mathbf{I_{\pm}}$
($\mathbf{k_{I}})$ and any of modes in $\mathbf{II}_{\pm}$ ($\mathbf{k_{II}}$),
which projects the state of atomic ensembles A, B' onto:
\begin{align}
|\psi_{\mathrm{ENG}}^{A,B'}\rangle & =\frac{1}{\sqrt{2}}(\hat{S}_{A,H,\mathbf{k_{I}}}^{\dagger}+\exp(i\phi)\hat{S}_{B',V,\mathbf{k_{I}}}^{\dagger})\label{eq:eng_state}\\
 & \times(\hat{S}_{A,V,\mathbf{k_{II}}}^{\dagger}+\exp(i\phi)\hat{S}_{B',H,\mathbf{k_{II}}}^{\dagger})|\text{vac}\rangle,
\end{align}
where $\phi$ gives the phase difference between optical paths from
$A$ and $B'$ to the midway CMS i.e. over distance $L_{0}/2$, and
$\hat{S}_{X,P,\mathbf{k}}^{\dagger}$ is the creation operator for
a spin-wave in ensemble $X$, in mode $\mathbf{k}$, with the corresponding
read-out photon $P$-polarized. Here and henceforth the modes are
indexed by the wavevector of the write-out photon $\mathbf{k}$ which
implicitly corresponds to a spin-wave in mode $\mathbf{K}$, which
upon readout generates a read-out photon in mode $\mathbf{-k}$. We
note here that since the coincidence can occur between any two modes
there is $M^{2}$ possibilities, in contrast to parallel scheme which
attempts ENG only between the $j$-th mode from $A$ and $j$-th mode
from $B'$.

The excitation number (DLCZ-type) entanglement present in $|\psi_{\mathrm{ENG}}^{A,B'}\rangle$
will be further converted to polarization entanglement during entanglement
connection (ENC) stage which also projects the state onto a subspace
containing one excitation per node.

Let us consider the probability of a successful ENG between $A$ and
$B'$ with multiplexing and by utilizing the memory modes in parallel.
If we denote the total probability of emitting two photons, transmission
and correct measurement outcome by $p_{1}$, then with $M$ modes
the total probability of successful ENG in any pair of modes in \emph{parallel}
operation would read
\begin{equation}
p_{g}^{(\mathrm{parallel})}=1-(1-p_{1})^{M}.\label{eq:peng_par}
\end{equation}
With multiplexing we harness the multi-mode single-photon detection
at CMS to provide the which-mode information about photons from each
memory, which further enables reconfiguration of a $M$-to-$1$ switch
at the memories' outputs. This way, we can entangle each of $M$ modes
from $A$ with each of $M$ modes from $B'$, effectively enhancing
the probability of entanglement generation 
\begin{equation}
p_{g}=1-(1-p_{1})^{M^{2}}.\label{eq:peng_mux}
\end{equation}
We note that this probability is equivalent to the parallel case with
$M^{2}$ modes.

Importantly, the single-mode ENG probability $p_{1}$ involves a fundamental
transmission loss which can be mitigated by $M^{2}$ scaling in $p_{g}$.
Wavevector multiplexed quantum memories can achieve around $5\times10^{3}$
modes, making the entanglement generation quasi-deterministic, even
with significant optical attenuation at large elementary distances
$L_{0}$.

\subsubsection*{Entanglement connection (ENC)}

Let us assume that ENGs have been successful between A and B', and
between B and C', to form a state $|\psi_{\mathrm{ENG}}^{A,B'}\rangle\varotimes|\psi_{\mathrm{ENG}}^{B,C'}\rangle$
involving modes $\mathbf{k_{I}}$, $\mathbf{k_{II}}$, $\mathbf{\mathbf{k_{III}}}$
and $\mathbf{\mathbf{k_{IV}}}$. Ensembles B and B' are at the same
node. We wish to carry ENC with B, B' so that A and C' share an entangled
state. Let us for the moment postpone the discussion of the multiplexing
stage and assume that read-out photons from B and B' occupying superpositions
of $\mathbf{k_{I}}$, $\mathbf{k_{II}}$ and $\mathbf{\mathbf{k_{III}}}$,
$\mathbf{\mathbf{k_{IV}}}$ modes, respectively, arrive at the ENC
segment in pre-selected photonic modes. For entanglement connection
(ENC) a two-photon interference is observed between the read-out photons
with single-photon detectors, as depicted in Fig. \ref{fig:engenc}b.
Once a coincidence with one photon per each depicted $\pm$ detector
is observed, the ENC succeeds with an output state depending on the
coincidence detectors signs $(\pm$,$\pm$). For example $(+,+)$
gives $(|HV\rangle_{A,C'}+|VH\rangle_{A,C'})/\sqrt{2}\equiv(\hat{S}_{A,H,\mathbf{k_{I}}}^{\dagger}\hat{S}_{C',V,\mathbf{\mathbf{k_{IV}}}}^{\dagger}+\hat{S}_{A,V,\mathbf{k_{II}}}^{\dagger}\hat{S}_{C',H,\mathbf{\mathbf{k_{III}}}}^{\dagger})/\sqrt{2}$,
while $(+,-)$ leads to $(|HH\rangle_{A,C'}+|VV\rangle_{A,C'})/\sqrt{2}$.
The output state can be corrected with a bit-flip operation so that
ENC always yields the same Bell state. While first-stage ENC (fENC)
taking place just after ENG requires additional half-wave plates in
the setup, further ENCs proceed without the second polarization rotation,
as depicted in Fig. \ref{fig:engenc}b, and require a phase-flip correction
instead of a bit-flip \citep{Jiang2007}.

\subsubsection*{Multiplexing (MUX)}

\begin{figure}
\includegraphics[width=1\columnwidth]{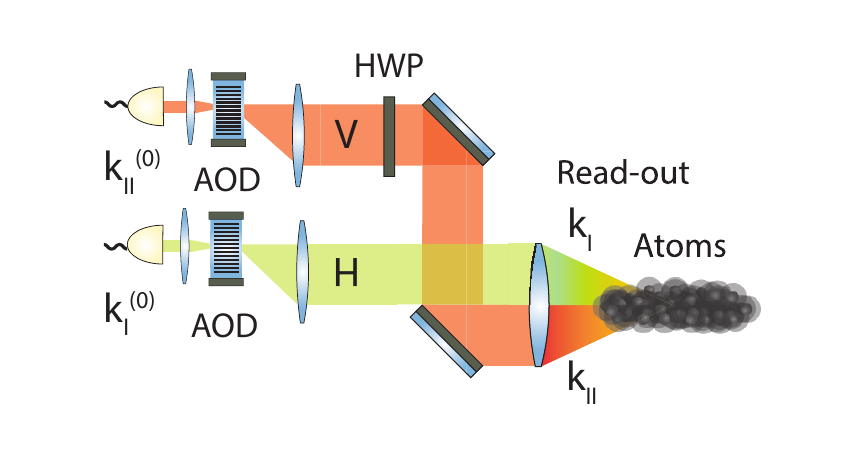}

\caption{Wavevector mode multiplexing with acousto-optical deflectors (AOD).
The which-mode information acquired during multimode entanglement
generation predicts the emission of a read-out photon at wavevector
modes $\mathbf{k_{I}}$ or $\mathbf{k_{II}}$, with the two modes
located in separate halves of the emission cone. The cone is split
in the far-field of the atomic cloud. Each half is multiplexed separately.
By imaging the atomic cloud onto the AOD we can shift the wavevector
of the read-out photon from $\mathbf{k_{I}}$ ($\mathbf{k_{II}}$)
to the desired mode $\mathbf{k_{I}^{(0)}}$ ($\mathbf{k_{II}^{(0)}}$)
which can be coupled to a single-mode fibre and routed to a single-mode
entanglement connection setup.\label{fig:mux_stark}}
\end{figure}

A critical step in the new protocol is multiplexing (MUX) the arbitrary
modes in $|\psi_{\mathrm{ENG}}\rangle$ to a pair of canonical pre-selected
modes, which enables two-photon interference between the read-out
photons, crucial for robust entanglement swapping \citep{Zhao2007,Jiang2007}.
Several strategies may be employed to implement the MUX stage. Importantly,
spatially-resolved detection involved in the new protocol at the ENG
station provides the required which-mode ($\mathbf{k_{I}}$, $\mathbf{k_{II}}$)
information. One idea may be to interface the memory with two read
beams at angles which are reconfigured via acousto-optical deflectors
(AOD) to facilitate the read-out at pre-select modes $\mathbf{\mathbf{k_{I}^{(0)}}},\mathbf{\mathbf{k_{II}^{(0)}}}$
\citep{Mazelanik2016}.

Another MUX method which may yield close to $100\%$ efficiency would
be to use fast acousto-optical deflectors (AOD) placed in the near-field
of the atomic cloud, to directly adjust the wavevector (angle) of
the read-out photons to match $\mathbf{k_{I}^{(0)}}$ and $\mathbf{\mathbf{k_{II}^{(0)}}}.$
The idea has been depicted in Fig. \ref{fig:mux_stark}. Importantly,
$\mathbf{k_{I}}$ and $\mathbf{k_{II}}$ always lay in different ($H$,
$V$) parts of the emission cone and thus can be separated in the
far-field and directed to two different AODs. Each AOD modulates the
read-out with only a single frequency reconfigured to match the difference
between the actual modes ($\mathbf{k_{I}}$, $\mathbf{\mathbf{k_{II}}}$)
and the target modes ($\mathbf{k_{I}^{(0)}}$, $\mathbf{k_{II}^{(0)}}$).

Alternatively, fast digital micromirror devices (DMDs) approaching
$\mu$s response times could be used in place of AODs offering even
broader possibilities \citep{Andersen2014}.

A more flexible technique could involve an in-memory MUX on the stored
spin-waves. The required technique – ac-Stark spin-wave modulation
– has been demonstrated \citep{Parniak2019,Mazelanik2019,Lipka2019}
in implementations of single–spin-wave inter-mode operations and spin-wave
wavevector ($\mathbf{K}$-space) displacements. In this approach,
the which-mode information is used to prepare a spin-wave $\mathbf{K}$-space
displacement operation which changes the wavevectors of stored spin-waves
to make the read-out photons match the pre-selected modes $\mathbf{\mathbf{k_{I}^{(0)}}}$,
$\mathbf{k_{II}^{(0)}}$.

\subsection*{Performance figure of merit}

\subsubsection*{Entanglement of formation ($E_{F}$)}

Inherently, due to the multi-excitation component in the generated
photon–spin-wave state, background noise and dark counts, part of
the detector counts will indicate randomly polarized photons. Therefore,
we model the experimental imperfections as a depolarizing channel,
obtaining in each memory mode a Bell Werner-like state $\rho_{|\psi\rangle}$
given by 
\begin{equation}
\rho_{|\psi\rangle}=\frac{(1-V)}{4}\hat{\mathbb{I}}+V|\psi\rangle\langle\psi|,
\end{equation}
where $|\psi\rangle$ is the selected Bell state $|\psi\rangle\in\{|\Psi_{\pm}\rangle,|\Phi_{\pm}\rangle\}$
and $V$ gives the interference visibility for a Bell-state measurement
(BSM). The first term $(1-V)/4\times\hat{\mathbb{I}}$ corresponds
to the introduced white noise.

The entanglement as a resource can be quantified with the entanglement
of distillation $E_{D}(\rho_{|\psi\rangle}^{\otimes n})=m/n$, which
gives the number $m$ of pure states $|\psi\rangle$ that can be distilled
from $n$ copies of $\rho_{|\psi\rangle}$ in the limit of $n\rightarrow\infty$.
In the opposite scenario the entanglement of formation $E_{F}(\rho_{|\psi\rangle}^{\otimes n})=m/n$
gives the required number $m$ of pure states $|\psi\rangle$ required
to create $n$ copies of $\rho_{|\psi\rangle}$. As entanglement of
distillation is generally difficult to calculate, we assume an optimistic
scenario and use entanglement of formation $E_{F}(\rho_{|\psi\rangle})\geq E_{D}(\rho_{|\psi\rangle})$
for the entangled bits (ebits) content of a single generated $\rho_{|\psi\rangle}$
state. For a Bell Werner-like state $\rho_{|\psi\rangle}(V)$ an exact
expression for $E_{F}(V)$ has been given in Refs. \citep{Wootters1998,Diaz-Solorzano2018}.
We note that other entanglement monotones are known \citep{Rains2001}
to give tighter bounds on $E_{D}$ yet for simplicity and taking into
account the estimative character of our calculations we use $E_{F}$.

\subsubsection*{Ebit rate per unit repeater cost ($Q$)}

The ebit rate $R$ quantifies the amount of distributed entanglement
bits per unit time. The average time $T_{\mathrm{tot}}$ between successful
entanglement distributions over the total distance $L$, gives the
protocol connection rate $1/T_{\mathrm{tot}}$ and the average entanglement
of formation $\langle E_{F}(V)\rangle$ of transmitted states yields
the ebit content per connection, thus the ebit rate is given by: 
\begin{equation}
R=\langle E_{F}(V)\rangle/T_{\mathrm{tot}}.\label{eq:eR}
\end{equation}
With a given total connection distance $L$, the number of employed
repeater nodes $N$ can be optimized to yield the highest ebit rate;
however, in design of practical repeater links the infrastructure
cost must be taken into consideration as well. To account for such
a limitation, we choose the ebit rate per unit repeater node cost
as our figure of merit: 
\begin{equation}
Q(N,L)=R(N,L)/N,\label{eq:Q}
\end{equation}
and optimize it over the number of nodes $N$ for each $L$: 
\begin{equation}
N^{\ast}(L)=\arg\max_{N}Q(N,L).\label{eq:Nstar}
\end{equation}

\subsection*{Quasi-deterministic entanglement generation}

\begin{figure}
\includegraphics[width=1\columnwidth]{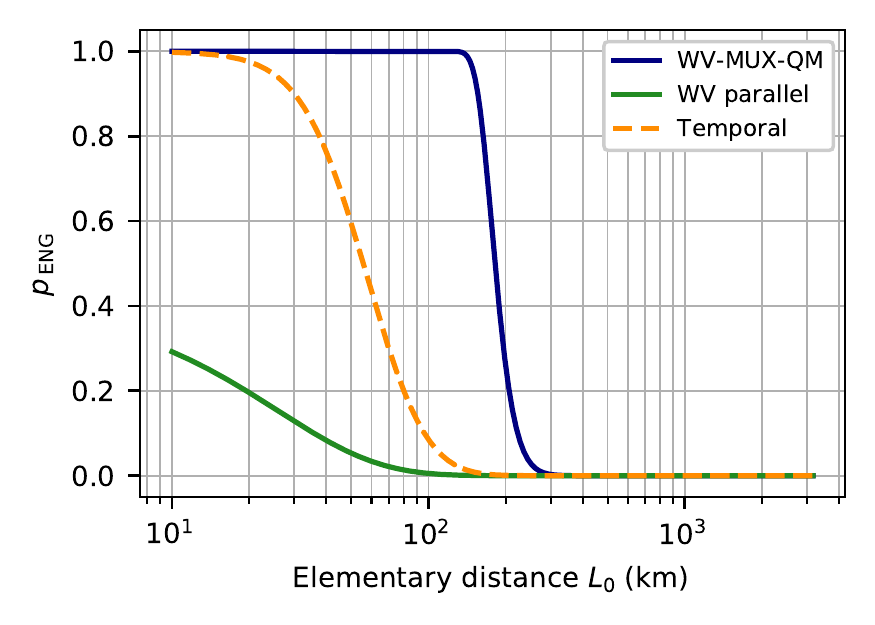}

\caption{Probability of entanglement generation in an elementary quantum repeater
link $p_{g}$ as a function of the inter-node distance $L_{0}$. Wavevector
multiplexed architecture (\emph{WV-MUX-QM}) is compared with a similar
wavevector multimode (\emph{WV parallel}) system without multiplexing,
and with a \emph{Temporal} multimode platform. See main text for model
parameters.\label{fig:peng}}
\end{figure}

Let us consider an exemplary link with $p_{1}=(\chi\eta_{m}\eta_{t})^{2},$where
$\eta_{m}$ denotes the multi-mode single-photon detection efficiency
and $\eta_{t}$ the transmission loss at $L_{0}/2$ distance between
a memory and the CMS. To keep the multi-excitation probability low,
the excitation probability is kept on the order of a few precents.
Here, we take $\chi=0.05$. Multimode detection compatible with thousands
of spatial modes can be realized with single-photon cameras e.g. a
CMOS camera with an image intensifier \citep{Lipka2018}, novel photon-counting
CMOS sensor \citep{Ma2017} or arrays of avalanche or superconducting
detectors \citep{Wollman2019}, or simply many detectors. The camera-based
solutions would be most suitable for free space transmission with
$800\;\text{nm}$ light, yielding at least $20\%$ detection efficiency.
On the other hand, in the telecom band one would have to use quantum
frequency conversion techniques (best overall efficiencies of the
order of $30\%$ \citep{Yu2020}) adapted for many modes along with
detector arrays. Overall, we will assume detection efficiency $\eta_{m}=0.2$
and consider telecom wavelength transmission over a fibre with attenuation
of around $\alpha=0.2\;\text{dB/\text{km}}$. For WV-MUX we will assume
$M=5500$ theoretically attainable in an experimentally demonstrated
system as detailed in \emph{Wavevector range }section. Figure \ref{fig:peng}
depicts the $p_{g}$ as a function of elementary distance $L_{0}$
for WV-MUX-QM-based link, compared to $p_{\mathrm{g}}^{(\mathrm{parallel})}$
for a wavevector multimode system where modes are connected in parallel
without multiplexing i.e. $i$-th mode of $A$ can be connected only
with $i$-th mode of $B$. Additionally, we consider \emph{temporal}
multimode system employing fast single-mode single-photon detection
with high detection efficiency $\eta_{s}\approx0.9$ and high photon
generation efficiency $\chi=0.$47 across $M=50$ temporal modes.
Noticeably, the entanglement generation is quasi-deterministic for
WV-MUX-QM system up to around $L_{0}\approx150\;\text{km}$.

\subsection*{Mode-specific entanglement quality}

\subsubsection*{Spin-wave decoherence}

The multi-photon contribution and noise can be quantified with the
second order intensity cross-correlation between the write-out and
read-out photons as given by:
\begin{equation}
g^{(2)}({\bf k_{w}},\mathbf{k_{r}})=\frac{\langle n_{w}({\bf k_{w}})n_{r}({\bf k_{r}})\rangle}{\langle n_{w}({\bf k_{w}})\rangle\langle n_{r}({\bf k_{r}})\rangle},\label{eq:g2}
\end{equation}
where $n_{w}({\bf k_{w}})$ ($n_{r}({\bf k_{r}})$) refers to the
number of write-out (read-out) photons with wavevector ${\bf k_{w}}$
(${\bf k_{r}}$) detected in a single experiment repetition, and the
average $\langle.\rangle$ is taken over the repetitions. With low
average photon numbers $n\ll1$ we can approximate $g^{(2)}(\mathbf{k_{w}},\mathbf{k_{r}})\approx p_{w,r}/(p_{w}p_{r})$,
where $p_{w,r}\equiv p_{w,r}(\mathbf{k_{w}},\mathbf{k_{r}})$ is the
total probability of observing a coincidence between write-out and
read-out photons with wavevectors $\mathbf{k_{w}}$and $\mathbf{k_{r}}$,
respectively, and $p_{w}\equiv p_{w}(\mathbf{k_{w}})$ ($p_{r}\equiv p_{r}(\mathbf{k_{r}})$)
gives the marginal probability of observing a write-out (read-out)
photon with wavevector $\mathbf{k_{w}}$ ($\mathbf{k_{r}}$). Coincidences
can be divided into uncorrelated events and those originating from
the correct memory operation $p_{w,r}=p_{w}p_{r}+p_{w,r}^{(\text{signal})}$.
Let us select the highest-correlated modes $\mathbf{k_{r}=-k_{w}=k}$.
If we denote the single-photon detection efficiency by $\eta$, read-out
efficiency by $\eta_{r}$ and the probability of a noise photon in
the read-out path as $B$, we get the following probabilities $p_{w,r}^{(\text{signal})}=\eta^{2}\eta_{r}\chi$,
$p_{w}=\eta\chi$, $p_{r}=\eta\eta_{r}\chi+B\eta$ which gives \citep{Zhao2009b}:
\begin{equation}
g^{(2)}(\mathbf{k},-\mathbf{k})\approx1+\left(\chi+\frac{B}{\eta_{r}}\right)^{-1}.\label{eq:g2beta}
\end{equation}
In general, noise and read-out efficiency depend on the memory storage
duration, subsequently deteriorating $g^{(2)}$. For clarity, we shall
include the $B/\eta_{r}$ term into a time-dependent effective $\tilde{\chi}(t)=\chi+B(t)/\eta_{r}(t)$.

\paragraph*{Visibility reduction from uncorrelated coincidences}

Among the registered coincidences between the write-out and read-out
photons there is a number of noise pairs originating from dark counts,
losses and multi-photon excitations. Importantly, in WV-MUX-QM such
noise photons attain horizontal or vertical polarization equiprobably,
reducing the visibility parameter of the generated state $V(\mathbf{k})$.
To derive $V(\mathbf{k})$ explicitly let us consider the second order
cross-correlation, as given by Eq. (\ref{eq:g2}). For a maximally
correlated pair of write-out $\mathbf{k}_{w}=\mathbf{k}$ and read-out
$\mathbf{k}_{r}=-\mathbf{k}$ wavevectors and with low average photon
numbers $n\ll1$ we have $g^{(2)}(\mathbf{k},\mathbf{-k})\approx p_{w,r}/(p_{w}p_{r})$,.
where $p_{w,r}\equiv p_{w,r}(\mathbf{k})$ is the single-experiment
probability of observing a coincidence between a write-out and read-out
photon and $p_{w}\equiv p_{w}(\mathbf{k})$ ($p_{r}\equiv p_{r}(-\mathbf{k})$)
denotes the marginal probability of observing a write-out (read-out)
photon. The coincidence probability can be written as $p_{w,r}=g^{(2)}p_{w}p_{r}$,
with $g^{(2)}\equiv g^{(2)}(\mathbf{k},-\mathbf{k})$. The state visibility
can be measured in a Bell state measurement (BSM) configuration. Consider
coincidence counts between write-out and read-out photons with a polarization
BSM. We can denote $V(\mathbf{k})=(p_{w,r}^{(+)}-p_{w,r}^{(-)})/(p_{w,r}^{(+)}+p_{w,r}^{(-)}),$where
superscripts $(\pm)$ denote measurements settings with maximally
constructive $(+)$ or destructive $(-)$ interference. In $(-)$
settings only noise coincidences are registered i.e. $p_{w,r}^{(-)}=p_{w}p_{r}$
while $p_{w,r}^{(+)}=p_{w,r}=g^{(2)}p_{w}p_{r}$. Therefore, the visibility
can be written as
\begin{equation}
V(\mathbf{k})=\frac{g^{(2)}(\mathbf{k},-\mathbf{k})-1}{g^{(2)}(\mathbf{k},-\mathbf{k})+1}.\label{eq:vtilde}
\end{equation}
We note that a $\mathbf{k}$ mode has a finite extent in $\mathbf{k}$-space
and for an efficient implementation of a BSM, it is necessary to integrate
the coincidences over the mode extent, effectively averaging the visibility,
given by Eq. (\ref{eq:vtilde}), in $\mathbf{k}$-space.

\subsubsection*{Memory decoherence\label{subsec:Memory-decoherence}}

The memory lifetime $\tau$, limited by the spin-wave decoherence
rate, sets a relationship between the ebit content of the generated
state $E_{F}$ and the storage time $t$. For an atomic ensemble cooled
in a magneto-optical trap (MOT) and stored in a dipole trap, the main
spin-wave decoherence mechanism is through the thermal motion of individual
atoms \citep{Parniak2017,Zhao2009b} which distorts the spatial structure
of a spin-wave. Intuitively, an atom travelling out of its initial
position has the more detrimental effect on the spin-wave, the finer
spatial details of the spin-wave are. Therefore, we expect $\tau$
to be mode-specific and grow with lower spin-wave wavevector modulus
$K\equiv|\mathbf{K}|$. The exact result follows from a thermal evolution
of a spin-wave state and is given by the overlap of the evolved and
the initial spin-wave states $|\langle S_{\mathbf{K}}(0)|S_{\mathbf{K}}(t)\rangle|^{2}\propto\exp(-t^{2}/\tau(K)^{2})$
which has a Gaussian form and where $\tau(K)=\gamma/K$ with the proportionality
constant $\gamma=\sqrt{m/k_{B}T}$ depending on the atomic mass of
Rb-87 $m$, ensemble temperature $T$ and Boltzmann constant $k_{B}$.
For typically attainable temperatures $T\approx1\;\text{\ensuremath{\mu}K}$,
$\gamma\approx10^{5}\;\mu\text{s}/\text{mm}$.

Effectively, the spin-wave decoherence decreases the readout efficiency
$\eta_{r}(t)=\exp(-t^{2}/\tau(K)^{2})$. We shall further assume a
constant noise level $B(t)=\text{const.}$ This way, we can account
for the visibility deterioration by considering the cross-correlation
$g^{(2)}(\mathbf{k},-\mathbf{k};\;t)\approx1+\exp(-t^{2}/\tau(K)^{2})/\tilde{\chi}(0)$
and using Eq. (\ref{eq:vtilde}) to arrive at the time-dependent average
visibility:
\begin{equation}
V(K,t)=\frac{1}{1+2\tilde{\chi}(0)\exp(t^{2}/\tau(K)^{2})},\label{eq:vtilde_t}
\end{equation}
where we shall approximate $\tilde{\chi}(0)\approx\chi$ for experimentally
feasible conditions.

\subsubsection*{Wavevector range}

Let us consider modes from $K_{\min}=10\;\text{mm}^{-1}$ to $K_{\max}=10^{3}\;\text{mm}^{-1}$
which constitute a practically feasible range of captured emission
angles $2\times0.073^{\circ}$ to $2\times7.3^{\circ}$ while still
allowing to route the write and read beams. In such a case we get
$\tau(K_{\min})=10\;\text{ms}$ and $\tau(K_{\max})=100\;\mu\text{s}$.
Importantly, the number of modes in a range $[K,K+\mathrm{d}K]$ is
$2\pi K\beta\mathrm{d}K$ which grows with $K$, and where $\beta$
is the $K$-space mode density. In our previous work \citep{Parniak2017}\emph{
}we have determined the number of modes in a WV-MUX-QM by performing
a singular value decomposition on experimentally collected data. The
obtained mode density $\beta=3.5\times10^{-3}\;1/\text{mm}^{-2}$
with our choice of $K_{\min}$ and $K_{\max}$ corresponds to the
total number of $M=5500$ modes, where implicitly we have halved the
number of modes to enable the generation of polarization entangled
states in $M$ pairs of wavevector modes, required for a two-photon
quantum repeater protocol.

The range of $K$ modes is limited by the geometry of the optical
setup and the size and resolution of the image sensor in a single-photon
sensitive camera \citep{Lipka2018}. The atomic ensembles is assumed
to be cooled in a magneto-optical trap and further trapped in a dipole
trap, reaching the temperature of a 1$\;\mu\text{K}$ which gives
$\gamma\approx10^{5}\;\mu\text{s}/\text{mm}^{-1}$ amounting to a
relatively short lifetime of around $100\;\mu\text{s}$ for $K_{\text{max}}$
mode. Clearly, there is little point in observing quickly decaying
modes for $K>K_{\text{\ensuremath{\max}}}$. The operation of the
memory requires counter-propagating write and read beams \citep{Parniak2017},
limiting the observable emission at small angles. With $800\;\text{nm}$
wavelength, $K_{\min}=10\;\text{mm}^{-1}$ amounts to ca. $200\;\mu\text{m}$
diameter in the far-field left for write/read beam routing, assuming
a lens of focal lengths $f=100\;\text{mm}$. Such a configuration
gives the diameter of the total observed far-field of around $1$
inch which is compatible with standard optical elements and feasible
to be imaged onto a commercially available $10\times10\;\text{mm}$
CMOS sensors with around $1\;\text{px}=10\;\mu\text{m}$ pixel size.
With a properly adjusted magnification the characteristic Gaussian
mode size $2\sigma$ of around $2\times4.8\;\text{mm}^{-1}$ corresponds
to around 2.5 px on the camera sensor \citep{Parniak2017}.

\subsubsection*{Mode performance}

\begin{figure}
\includegraphics[width=1\columnwidth]{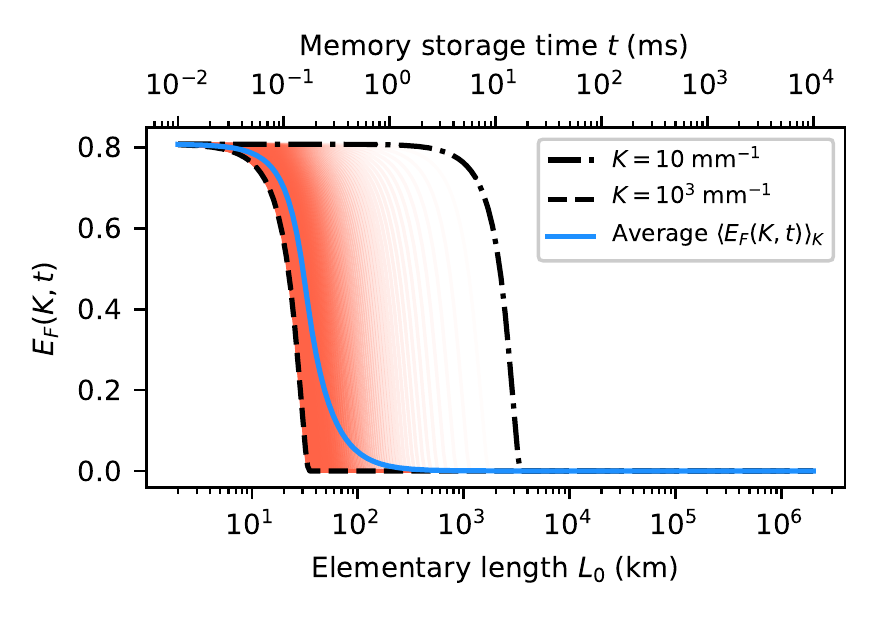}

\caption{Entanglement of formation $E_{F}(K,L_{0}/c)$ as a measure of ebit
content in entangled states of a photon and a spin-wave with wavevector
modulus $K$, stored for a memory time $L_{0}/c$ required in entanglement
generation over an inter-node elementary distance $L_{0}$. Red curves
correspond to specific memory modes while the blue curve gives the
average over the modes. The black dashed and dot-dashed curves indicate
the $E_{F}(K_{\max},t)$ and $E_{F}(K_{\min},t)$, respectively, with
$K_{\max}=10^{3}\;\text{mm}^{-1}$, $K_{\min}=10\;\text{mm}^{-1}$
corresponding to the modes with the highest and lowest spin-wave wavevector.
\label{fig:eof_kt}}
\end{figure}

The ebit content of states generated across different $K$ modes varies,
as illustrated in Fig. \ref{fig:eof_kt}. As the number of modes with
a given $K$ grow with $K$, most of the modes occupy quickly decaying
high-$K$ modes, reducing the average entanglement of formation $\langle E_{F}(K,t)\rangle_{K}$.

\subsection*{Quantum repeaters architecture}

\subsubsection*{Hierarchy}

Hierarchical protocols like DLCZ or two-photon protocols divide the
$N$ nodes pair-wise recursively to form a binary tree. There is a
conceptual advantage in a such an approach as one only needs to consider
the connection time between two sub-nets at each nesting level; however,
if one sub-net fails and needs to start from the entanglement generation
at the lowest level, there is a tremendous waiting time overhead for
classical communication to synchronize the manoeuvre. The average
requirement for memory storage time may many times exceeds the direct
communication time $L/c$ between the parties.

On the other hand, in ahierarchical architecture all nodes would operate
synchronized only by a classical clock with period $L_{0}/c$ and
without any feedback. While the requirements for memory storage time
are substantially lower, all nodes must succeed simultaneously for
an overall success. With limited efficiencies of memories and detectors,
successful generation becomes exponentially difficult with the increasing
number of nodes.

Semihierarchical architecture has been proposed \citep{Liu2017} as
an intermediate regime. Nodes communicate with a central station located
mid-way between the final parties, which waits until all nodes successfully
generate the entanglement and synchronizes entanglement connection.
While the ENC has to succeed simultaneously across all nodes, the
nodes which succeed with ENG wait for other nodes. Such an approach
improves the probability of success per protocol repetition as compared
with ahierarchical approach and has lower memory time requirements
than hierarchical architecture.

Regardless of the architecture, the average time for successful entanglement
distribution can be modularly written as
\begin{equation}
T_{\mathrm{tot}}=T_{r}/(P_{\mathrm{ENG}}P_{\mathrm{ENC}}\eta_{s}^{2}\eta_{x}^{2}),\label{eq:rate}
\end{equation}
where $T_{r}$ is the (average) repetition period of the protocol,
$P_{\mathrm{ENG}}$ the probability of a successful entanglement generation
between all pairs of ensembles, $P_{\mathrm{ENC}}$ the probability
of a successful entanglement connection across all nodes and $\eta_{s}^{2}\eta_{x}^{2}$
the probability of detecting the entangled photons at final parties
with $\eta_{s}$ being the efficiency of single-photon detectors and
$\eta_{x}$ of multiplexing.

\subsubsection*{Ahierarchical architecture}

Let us assume that the main delay in the protocol is the transmission
time of a write-out photon between a node and a CMS and of the which-mode
information back to the node i.e. $T_{r}=L_{0}/c$, where we take
$c=0.2\;\text{km}/\mu$s for a fibre transmission. For the protocol
to succeed with $N=L/L_{0}+1$ nodes, we need to generate entanglement
(ENG) between $N-1$ memories and perform first-stage ENC (further
ENC) between $\approx\lceil(N-2)/2\rceil$ (between $\lfloor(N-2)/2\rfloor$)
memories. Additionally multiplexing (MUX) is required before each
ENC and fist-stage ENC (fENC). Let us denote the efficiencies of ENG
in a single mode, fENC, ENC and MUX by $p_{g}$, $p_{f}$, $p_{e}$,
$\eta_{x}$, respectively. In the ahierarchical architecture all nodes
blindly assume that all other nodes succeeded with ENG and proceed
with the ENC. The overall probability for ENG is given by 
\begin{equation}
P_{\mathrm{ENG}}=p_{g}^{N-1},
\end{equation}
with $p_{g}$ given by Eq. (\ref{eq:peng_mux}), while for ENC it
is 
\begin{equation}
P_{\mathrm{ENC}}=p_{f}^{\lceil(N-2)/2\rceil}p_{e}^{\lfloor(N-2)/2\rfloor}\eta_{x}^{N}.
\end{equation}
In standard two-photon protocols $p_{e}$ is upper bounded by $1/2$
and $p_{f}=p_{e}/4$ due to post-selecting on coincidence patterns
chosen to project the state of connected memories onto a Bell state
\citep{Zhao2007}.

\subsubsection*{Semihierarchical architecture}

For the rates in semihierarchical architecture we follow the derivations
by Liu \emph{et al.} \citep{Liu2017} with a slight modification to
match the two-photon protocol and the WV-MUX-QM platform. Let us start
with Eq. (\ref{eq:rate}). We shall modify the $T_{r}/P_{\mathrm{ENG}}$
factor to be now an expectation value over the distribution of waiting
times for all nodes to accomplish entanglement generation. Importantly,
there is an additional waiting overhead $L/c$ for two-way communication
with the central station. The factor $T_{r}/P_{\mathrm{ENG}}$ now
reads $T_{r}/P_{\mathrm{ENG}}=(L_{0}/c\times f(N,p_{g})/p_{g}+L/c)$
where $f(N,p_{g})$ gives the expected number of ENG repetitions to
accomplish ENG between $N$ nodes with $p_{g}$ given by Eq. (\ref{eq:peng_mux})
or Eq. (\ref{eq:peng_par}) for multiplexed and parallel platforms,
respectively. The factor $f(N,p_{g})$ is given by:
\begin{equation}
f(N,p_{g})=p_{g}\sum_{j=1}^{\infty}j\times\{[1-(1-p_{g})^{j}]^{N}\}-[1-(1-p_{g})^{j-1}]^{N}\}.
\end{equation}

\subsection*{Limitations}

\subsubsection*{Maximal range}

A non-zero ebit content $E_{F}(V(t_{m}))>0$ of the generated state
after storage time $t_{m}$ requires visibility $V(t_{m})$ above
$1/3$ which is equivalent to $\chi\exp[t_{m}^{2}/\tau(K)^{2}]<1$
by Eq. (\ref{eq:vtilde_t}).

In the case of ahierarchical architecture the storage time is always
$t_{m}=L_{0}/c$ giving $L_{0}^{(\max)}=c\tau(K)\sqrt{\log(1/\chi)}$
and so $L_{\max}=NL_{0}^{(\max)}$. For $\chi=0.05$ and $c=0.2\;\text{km}/\mu\text{s}$
we have for the longest-live $K=10\;\text{mm}^{-1}$ mode $\tau(K)=10\;\text{ms}$
and $L_{0}^{(\max)}\approx3.5\times10^{3}\;\text{km}$. Let us require
that $1\%$ of all modes have a non-zero ebit content, this corresponds
to a range of $K$ from $10\;\text{mm}^{-1}$ to $100\;\text{mm}^{-1}$
with the lowest $\tau(K=100\;\text{mm}^{-1})=1\;\text{ms}$ and in
this case $L_{0}^{(\max)}\approx350\;\text{km}$. This characteristic
range is clearly visible in Fig. \ref{fig:eof_kt}.

For the semihierarchical architecture and assuming all nodes succeed
with entanglement generation in the first try, the total storage time
for each memory is $t_{m}=(L+L_{0})/c$. Employing Eq. (\ref{eq:vtilde_t}),
the requirement on the memory coherence time becomes:

\begin{equation}
\tau(K)>2\frac{L+L_{0}}{c\log(1/\chi)},
\end{equation}
which for a fixed $\tau(K)$ sets the maximal range
\begin{equation}
L_{\max}=\frac{N-1}{N}\frac{\tau(K)}{2}c\log\left(\frac{1}{\chi}\right),
\end{equation}
where $L+L_{0}=LN/(N-1)$. For $\chi=0.05$, $\tau(K)=10\;\text{ms}$,
$c=0.2\;\text{km}/\text{\ensuremath{\mu}s}$, the maximal distance
is ca. $L_{\max}\approx3\times10^{3}\;\text{km}$ for large $N\gg1$.

\subsubsection*{Entanglement connection probability}

\begin{figure}
\includegraphics[width=1\columnwidth]{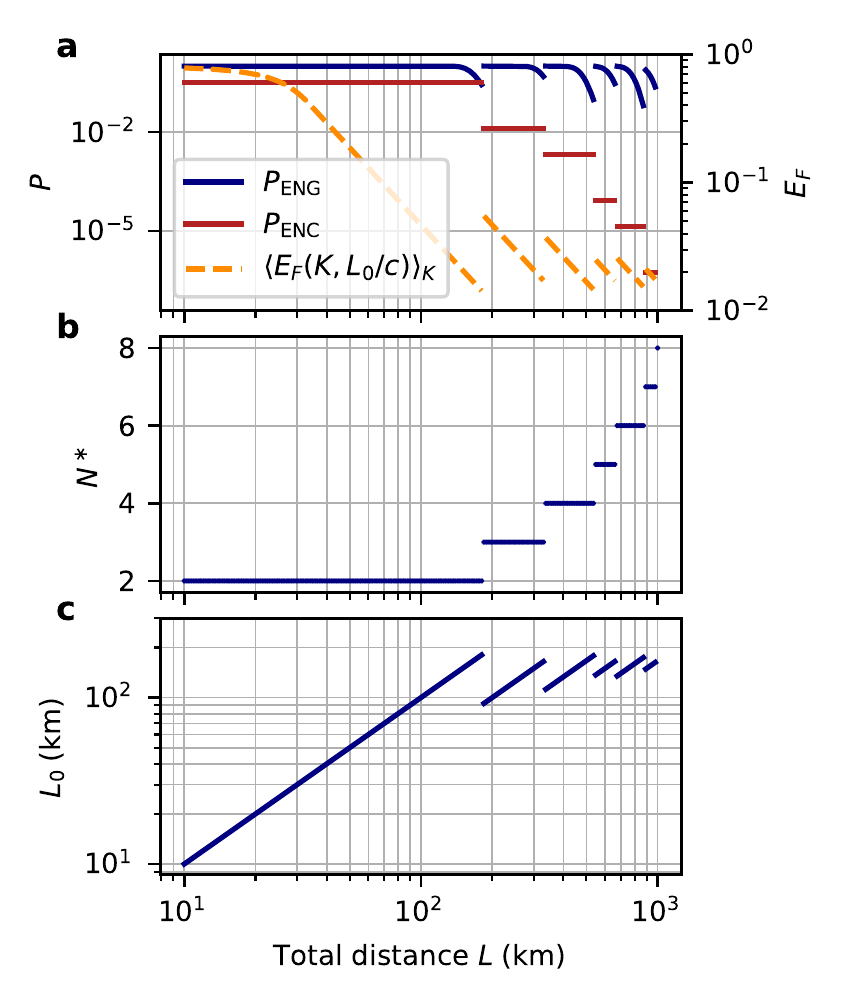}

\caption{Performance measures for WV-MUX-QM platform in ahierarchical architecture
with a total distance $L$.\textbf{ a} Total probability of entanglement
generation in all nodes simultaneously $P_{\mathrm{ENG}}$ is nearly
constant and close to unity while total probability of entanglement
connection between all nodes and final detection of entangled photons
$P_{\mathrm{ENC}}$ falls significantly as the optimal total number
of nodes $N^{\ast}$ – depicted in \textbf{b} – grows.\textbf{ c}
Elementary distance between adjacent nodes $L_{0}$ for the optimal
number of nodes $N^{\ast}$.\label{fig:probs}}
\end{figure}

The tremendous number of modes in WV-MUX-QM platform combined with
flexible multiplexing offers a quasi-deterministic entanglement generation
for extended distances between the elementary nodes. As depicted in
Fig. \ref{fig:probs}a, the probability of a successful ENG between
any of $M^{2}$ combinations of modes remains nearly $100\%$ even
with significant optical losses ($1\%$ transmission over $100\;\text{km}$).
Importantly, lifetime of spin-waves occupying different $K$ modes
quickly deteriorates with increasing $K$, making only a small fraction
of modes feasible for ENG over large $L_{0}$, as indicated by the
average entanglement of formation $\langle E_{F}(K,L_{0}/c)\rangle_{K}$
which is depicted in Fig. \ref{fig:probs}a. This in turn limits the
maximal $L_{0}$ which settles at the level of around $150\;\text{km}$,
as illustrated in Fig. \ref{fig:probs}c. With increasing $L$ and
limited $L_{0}$ the optimal number of nodes $N^{\ast}$ quickly increases,
which is clearly visible in Fig. \ref{fig:probs}b. Conversely, as
indicated by Eq. (\ref{eq:rate}), the probability of a simultaneous
success of ENC across all nodes, scaling with the power of $N^{\ast}$,
rapidly decreases.

A possible solution would be to entangle many pairs of modes during
ENG and perform a multiplexed ENC amending the deteriorating scaling
with the power of $N^{\ast}$.

\subsubsection*{Telecom wavelength and multimode transmission}

While we start our discussion from an experimental realization of
a WV-MUX-QM platform \citep{Parniak2017} which employs a $\approx800\;\text{nm}$
light-matter interface, there have been several demonstrations of
Rb-87 atomic memories working in the telecom regime thanks to external
or in-memory conversion \citep{Albrecht2014,Chaneliere2006,Chang2019,Radnaev2010};
therefore, an experimental realization could build on these results
and exploit wavevector multiplexing together with a telecom light-matter
interface.

Another technically challenging task would be to implement multimode
transmission channel. One way would be to use free-space which is
inherently multimode, yet sensitive to atmospheric conditions and
requiring expensive optical infrastructure. Another solution are multimode
fiber transmission systems e.g. consisting of an array of single-mode
telecom \foreignlanguage{american}{fibers}. Importantly, commercially
available fiber coupled microlens arrays \citep{SQSVlaknovaoptikaa.s}
enable efficient coupling of particular memory modes to the transmission
channel and make the solution practically feasible.

\subsection*{Rate comparison}

We have compared the performance of several physical platforms as
candidates for a quantum repeater nodes in ahierarchical and semihierarchical
architecture. For each total distance between the parties $L$ we
select a number of nodes $N\geq2$ which maximizes the entanglement
transfer rate per node number given by Eq. \eqref{eq:Q}: $N^{\ast}(L)=\arg\max_{N}\langle Q(N,L)\rangle_{K,t}$
with the average $\langle.\rangle$ taken over the distribution of
required memory times $t$ across realizations and the performance
of individual $K$-modes in a single realization. The comparison encompasses
the wavevector multiplexed quantum memory (\emph{WV-MUX-QM}), parallel
operation (\emph{parallel}, without multiplexing) of a wavevector
multimode quantum memory, temporal multimode quantum memory (\emph{temporal})
and a state-of-the-art single-mode, long-lifetime atomic quantum memory
in an optical lattice (\emph{lattice}). For reference we consider
a direct entanglement generation via a single spontaneous parametric
down conversion (SPDC) source located midway between the parties.
\begin{figure}
\includegraphics[width=1\columnwidth]{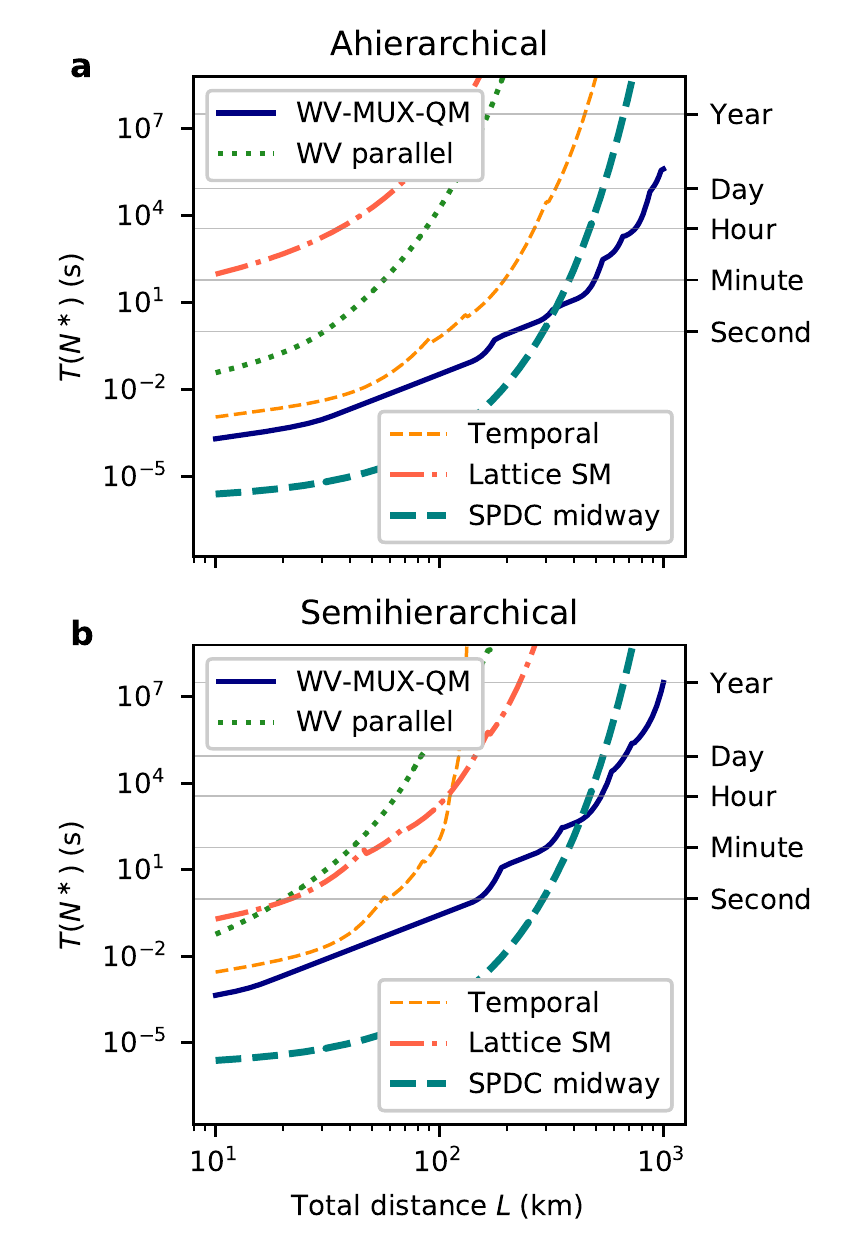}

\caption{Average time $T$ for a successful transfer of an ebit between parties
separated by distance $L$, employing either a central SPDC source
as a generator of polarization-entangled pairs of photons (\emph{SPDC})
or a quantum repeater architecture (\emph{WV-MUX-QM, WV parallel,
Temporal, Lattice} \emph{SM}), either \textbf{a} - ahierarchical or
\textbf{b} - semihierarchical. Experimentally feasible physical realizations
of a repeater node are compared. The number of repeater nodes $N$
is optimized for a maximal ebit rate per number of nodes $Q$.\label{fig:rate}}
\end{figure}

Figure \ref{fig:rate} depicts the average time for a successful transfer
and detection of a single ebit $T=1/(N^{\ast}\times Q(N^{\ast},L))$
for the compared platforms, along typical time scales. Above ca. $300\;\text{km}$
WV-MUX-QM in the ahierarchical architecture outperforms a direct SPDC
source. For ca. $550\;\text{km}$ the average time for the WV-MUX-QM
platform is ca. 6 minutes while for an SPDC source it amounts to over
$2$ days; for ca. $700\;\text{km}$ the average times are around
$40$ minutes versus over $5$ years.

Importantly, the assumed parameters correspond to the state-of-the-art
experimentally attainable systems; therefore, while the current ebit
rates remain impractical for most technological applications, wavevector
multiplexing appears to be a promising technique for near-term quantum
repeaters.

\begin{figure}
\includegraphics[width=1\columnwidth]{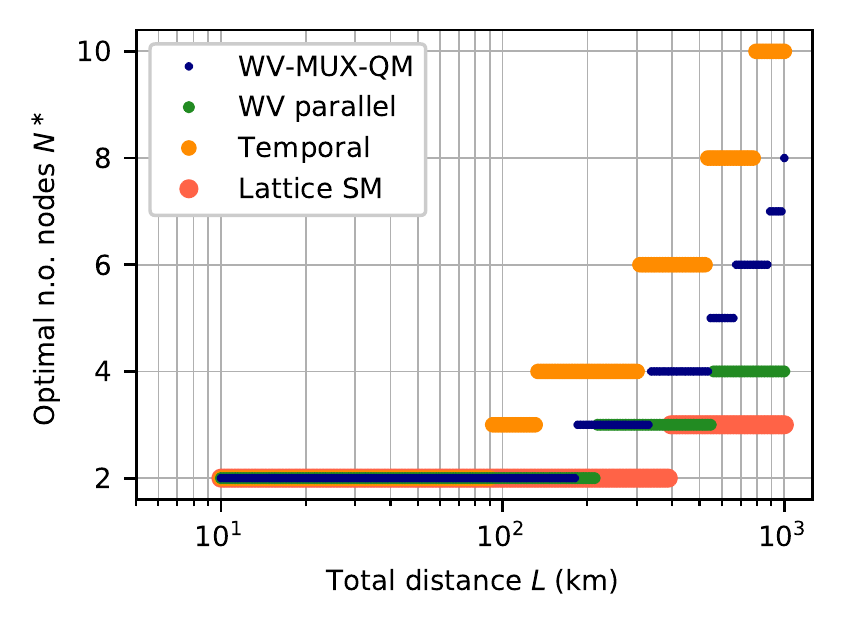}

\caption{Optimal number of repeater nodes $N^{\ast}$ maximizing the ebit rate
per number of nodes $Q$ for a given total distance $L$. Physical
platforms for the repeater node are compared within the ahierarchical
architecture.\label{fig:nopt}}
\end{figure}

As illustrated in Fig. \ref{fig:nopt}, the optimal number of nodes
$N^{\ast}(L)$ grows relatively slowly with $L$ for the WV-MUX-QM
platform, reflecting the ability to quasi-deterministically generate
entanglement over extended inter-node elementary distances $L_{0}$.

One of the limiting factors for the WV-MUX-QM architecture is the
mode-dependent memory lifetime which quickly degrades the ebit content
of a significant number of modes. For example, for an elementary distance
of $L_{0}=150\;\text{km}$, the state generation is still quasi-deterministic,
while the entanglement of formation yields on average $\langle E_{F}(K,t=L_{0}/c)\rangle_{K}\approx2\times10^{-2}$.
Importantly, despite very low ebit content per state, the are probabilistic
protocols which enable single-copy distillation of a pure Bell state
\citep{Wang2006}.

\section{Discussion}

In this work we show the feasibility of wavevector multiplexed quantum
memories (WV-MUX-QM) for near-term quantum repeaters, as an alternative
to temporally, spectrally or spatially-multimode platforms. Remaining
within the constraints of current technology, we consider an extension
of an experimentally demonstrated WV-MUX-QM platform \citep{Parniak2017}
combined with available methods of optical multiplexing and fast spatially-resolved
single-photon detection. We harness the tremendous number of modes
$M\approx5500$ theoretically available in WV-MUX-QM to design a robust
scheme for polarization entanglement generation (ENG) between distant
quantum repeater nodes. Importantly, the designed ENG scheme allows
to connect any pair of modes giving $M^{2}$ possibilities and providing
a quasi-deterministic ENG at ca. $150\;\text{km}$ of telecom fibre.

We extend the WV-MUX-QM setup with a specifically designed Mach-Zehnder
interferometer allowing implementation of a two-photon interference-based
quantum repeater protocol \citep{Zhao2007} robust to optical phase
fluctuations which deteriorate the performance of the standard DLCZ
protocol. Additionally, we consider each memory mode separately by
employing a mode-specific decoherence model and calculate the ebit
transfer rate by modelling the losses and noise as an depolarizing
channel. For the resource-efficient comparison of different quantum
repeater platforms, we establish an ebit transfer rate between the
final parties per number of employed quantum repeater nodes as our
figure of merit.

Finally, we analyse WV-MUX-QM in recently proposed semihierarchical
quantum repeater architecture as well as a simple ahierarchical architecture
and find the main limitations of the WV-MUX-QM platform. Importantly,
while the entanglement generation can be quasi-deterministic for over
a hundred km inter-node distance $L_{0}$, most of the memory modes
have a short lifetime deteriorating the ebit content of the generated
states with an increasing $L_{0}$. Conversely, the total distance
$L$ is mainly limited by the growing number of nodes $N$ with a
limited inter-node distance $L_{0}\lessapprox150\;\text{km}$. Importantly,
while multiplexing remedies the fundamental transmission losses inherent
to the ENG stage, the entanglement connection (ENC) remains single-mode.
Since ENC results are post-selected, even with ideal detectors and
memory retrieval efficiency, the probability of a successful ENC is
severely limited. With the growing number of nodes $N$, the requirement
for a simultaneous success of all $N-1$ ENCs constitutes the main
factor limiting the total distance $L$. Furthermore, we envisage
that our methods can be also applied to a spatially-multiplexed quantum
memory \citep{Pu2017}, where multiplexing is provided by addressing
different micro-ensembled at different times with AODs. Since the
WV-MUX-QM platform is capable of flexible processing on the stored
spin-waves \citep{Parniak2019,Mazelanik2019,Lipka2019}, we envisage
that, as a remedy to the low total ENC probability, a method could
be design to entangle many mode pairs simultaneously during ENG and
perform a multiplexed or parallelized ENC, such an approach is however
beyond the scope of this work and may be an interesting avenue for
future development.

\section*{Author contributions}

M.L., M.M. and M.P. jointly conceived the scheme. M.L. developed the
theory, performed calculations and wrote the paper with contributions
from other authors. M.P. supervised the project.

\section*{Competing Interests}
The authors declare that there are no competing interests.

\section*{Data Availability}
Data generated during the study is available from the corresponding author upon reasonable request.

\begin{acknowledgments}
This work was funded by Ministry of Science and Higher Education (Poland)
grants No. DI2016 014846, DI2018 010848; National Science Centre (Poland)
grants No. 2016/21/B/ST2/02559, 2017/25/N/ST2/01163, 2017/25/N/ST2/00713;
Foundation for Polish Science MAB 4/2017 ``Quantum Optical Technologies''.

The \textquotedbl Quantum Optical Technologies ” project is carried
out within the International Research Agendas programme of the Foundation
for Polish Science co-financed by the European Union under the European
Regional Development Fund. M.P. was supported by the Foundation for
Polish Science START scholarship.

We would like to thank W. Wasilewski for fruitful discussions and
K. Banaszek for the generous support.
\end{acknowledgments}

\section*{Methods}

In our comparison we assume that the transmitted ebit has to be detected
at final parties to be useful i.e. we include the efficiency of detectors
at the first and the last node. The optical transmission is assumed
to be carried via telecom fibres characterized by attenuation of $\alpha=0.2\;\text{dB}/\text{km}$
which gives the transmission efficiency over length $z$ of $\eta_{t}(z)=10^{-\alpha z/10}$.
The success probability for a single ENG is taken as $p_{1}=(\chi\eta_{t}(L_{0}/2)\text{\ensuremath{\eta_{m}}})^{2}$,
while for ENC to be $p_{e}=(\eta_{r}\eta_{s})^{2}/2$. Whenever we
are detecting single-photons in a single spatial mode, we assume the
detector efficiency of $\eta_{s}=0.9$ \citep{Marsili2013} and for
the multimode detection $\eta_{m}=0.2$ \citep{Lipka2018}, both corresponding
to experimentally attainable values. Model parameters are summarized
in Tab. \ref{tab:pars}.

\begin{table}
\begin{tabular}{c|c|c|c|c|c|c|c}
\multicolumn{8}{c}{Model parameters}\tabularnewline
\hline 
\hline 
Platform & $M$ & $\chi$ & $\tau\;(\text{ms})$ & $\eta_{x}$ & $\eta_{r}$ & $\eta_{s}$ & $\eta_{m}$\tabularnewline
\hline 
\hline 
WV-MUX-QM & 5500 & $0.05$ & $(10^{2}\,\mathrm{{mm^{-1}}})/K$ & 0.9 & 0.7 & 0.9 & 0.2\tabularnewline
\hline 
WV parallel & 5500 & $0.05$ & $(10^{2}\,\mathrm{{mm^{-1}}})/K$ & 1 & 0.7 & 0.2 & 0.2\tabularnewline
\hline 
Temporal & 50 & $0.47$ & $1$ & 1 & 0.71 & 0.9 & 0.9\tabularnewline
\hline 
Lattice SM & 1 & $0.05$ & $220$ & 1 & 0.76 & 0.9 & 0.9\tabularnewline
\end{tabular}\caption{Summary of model parameters for the comparison of quantum repeater
platforms. Wavevector-multiplexed (WV-MUX-QM) and wavevector parallel
(WV parallel) atomic quantum memories are compared with solid-state
temporal multimode and single-mode state-of-the-art atomic memory.
Parameters are estimated from current state-of-the-art experimental
demonstrations.}
\label{tab:pars}
\end{table}

For the SPDC midway source we assume a repetition rate of $f_{\mathrm{rep}}=80\;\text{MHz}$,
the probability of generating a pair of correlated photons $\chi=0.01$
and perfect visibility $V=1$ giving $E_{F}=1$ i.e. one ebit per
successful connection. The average time for an ebit transmission is
therefore $T_{\text{SPDC}}(L)=\chi(\eta_{t}(L)\eta_{s})^{2}f_{\mathrm{rep}}$.

For the \emph{WV-MUX-QM} and \emph{WV parallel }models we assume an
atomic Rb-87 quantum memory with wavevector modes. The probability
of generating a pair of a photon and a spin-wave is assumed to be
$\chi=0.05$ which is found close to the optimal for long total distances
$L\geq100\;\text{km}$. The readout efficiency is assumed at the state-of-the-art
experimentally-demonstrated level of around 70\% \citep{Cho2016}
i.e. $\eta_{r}=0.7$.

For \emph{WV-MUX-QM }the multiplexing efficiency is taken to be $\eta_{x}=0.9$
which is a realistic estimate with an optical switch based on acousto-optical
deflectors, such as demonstrated in ref. \citep{Jiang2019} and detailed
in \emph{Multiplexing }section. The probability of entanglement generation
between two nodes is in this case given by Eq. (\ref{eq:peng_mux}).

For \emph{WV parallel} operation there is no loss at multiplexing
thus $\eta_{x}=1$, and the probability of entanglement, given by
Eq. \ref{eq:peng_par}, scales with $M$ and not $M^{2}$. Additionally,
lack of multiplexing requires a mode-resolved BSM for ENC stage, reducing
the detection efficiency at this stage from $\eta_{s}=0.9$ to $\eta_{m}=0.2$.

For \emph{Temporal} multiplexing we consider solid-state quantum memories
with separate single-photon sources (SPS) e.g. quantum dot sources
and memories exploiting atomic frequency comb in rare-earth-ion-doped
solids. While multiple degrees of freedom (DoF) multiplexing has been
demonstrated in such systems \citep{Yang2018}, there is no straight-forward
way to harness additional DoF to implement two-photon protocols without
the overhead of multiple pairs of SPS and QM at each node. Therefore,
we assume temporal modes are processed independently and Eq. (\ref{eq:peng_par})
applies. The assumed number of modes is as reported in the state-of-the-art
experimental demonstrations $M=50$ with a lifetime of $1\;\text{ms}$
for all modes \citep{Jobez2016}. For the assumed quantum dot SPS
a $66\%$ generation efficiency has been demonstrated \citep{Ding2016}.
The total memory storage and retrieval efficiency is taken to be $50\%$,
which is around the state-of-the-art results \citep{Sabooni2013},
and assumed to be equally distributed over the storage $\chi=0.66/\sqrt{2}\approx0.47$
and the retrieval $\eta_{r}=1/\sqrt{2}\approx0.71$ efficiencies.

We note that across various implementations of solid state quantum
memories in doped crystals, there is a high discrepancy between storage
times, efficiency, number of modes and multiplexing capabilities.
Instead of considering a single crystal or protocol, for the purpose
of this comparison, we optimistically assume a platform performing
reasonably close to all of the state-of-the-art parameters.

Finally, the \emph{Lattice SM} platform refers to a single mode (SM)
atomic memory with atoms trapped in an optical lattice which greatly
extends the memory lifetime to ca. $220\;\text{ms}$ as reported by
Yang \emph{et al. }\citep{Yang2016}. Simultaneous use of an optical
cavity in this experiment increased the readout efficiency to ca.
$76\%$, we thus assume $\eta_{r}=0.76$. Other parameters are taken
as in the \emph{parallel} case, except the number of modes $M=1$.

Importantly, for \emph{Lattice SM} and \emph{temporal }platforms,
the decoherence is exponential $\exp(-t/\tau)$, while for wavevector
multimode memories it has a Gaussian profile $\exp(-t^{2}/\tau^{2})$.

\bibliographystyle{apsrev4-1}
\bibliography{bell_theor}

\end{document}